**PAPER**

# Coarse-Grained Molecular Simulation of Extracellular Vesicles Squeezing for Drug Loading

Khayrul Islam [a], Meghdad Razizadeh [b] and Yaling Liu *[a,c]

In recent years, extracellular vesicles such have become promising carriers as the next-generation drug delivery platforms. Effective loading of exogenous cargos without compromising the extracellular vesicle membrane is a major challenge. Rapid squeezing through nanofluidic channels is a widely used approach to load exogenous cargos into the EV through the nanopores generated temporarily on the membrane. However, the exact mechanism and dynamics of nanopores opening, as well as cargo loading through nanopores during the squeezing process remains unknown and is impossible to be visualized or quantified experimentally due to the small size of the EV and the fast transient process. This paper developed a systemic algorithm to simulate nanopore formation and predict drug loading during extracellular vesicle (EV) squeezing by leveraging the power of coarse-grain (CG) molecular dynamics simulations with fluid dynamics. The EV CG beads are coupled with implicit Fluctuating Lattice Boltzmann solvent. Effects of EV property and various squeezing test parameters, such as EV size, flow velocity, channel width, and length, on pore formation and drug loading efficiency are analyzed. Based on the simulation results, a phase diagram is provided as a design guidance for nanochannel geometry and squeezing velocity to generate pores on membrane without damaging the EV. This method can be utilized to optimize the nanofluidic device configuration and flow setup to obtain desired drug loading into EVs.

## Introduction

Biomembranes are among the most crucial parts of every living organism. The complex structure of biomembranes is composed of various fluid-like amphiphilic phospholipid molecules, transmembrane proteins, and carbohydrates. Extracellular vesicle contains fluid-like closed structures of phospholipids suspended in solutions and have wide application in biophysics and nanomedicine due to their similarity with natural biomembranes. For instance, EVs are widely used as drug delivery vehicles. [1,2] Being naturally derived composition and function as intracellular communication tools, they are free from several disadvantages compared to liposomal drug delivery, *e.g.*, endosomal degradation, immune clearance, organ toxicity, and insertional mutagenesis.[3–6] Commonly used drug loading methods for EVs include passive incubation and electroporation. Passive incubation has the drawback of long incubation time and low loading efficiency [7–9] where electroporation can cause significant damage to both EV and their cargos. [7,8,10,11] Alternatively, nanofluidic squeezing can easily manipulate flow and samples of interest, thus showing great potential for EV drug loading. [12,13] The process includes mixing EV and drug suspension, injection through a microfluidic device, and squeezing EVs through nanochannel to generate transient nanopores on the membrane, thus allowing targeted cargos to enter the EV. Transporting cargo through the membrane is a diffusion process that continues until the pore heals if the saturation level is not reached.

Various experimental methods have been developed to study the diffusion of lipid bilayers [14], deformations of vesicles [15,16], and membrane fusion. [17] For instance, super-resolution imaging techniques have progressed significantly to investigate the dynamics of biological membranes. [18,19] However, the pore formation and rupture of EVs under large deformations are hard to be studied with experimental approaches due to their small scale and transient short time scales. The current experimental setup and testing conditions are still based on trial and error and are far from achieving the best loading performance. Various numerical approaches have been developed during the last three decades to complement experimental understandings of biomembranes. While All Atomic (AA) modeling techniques are still progressing in terms of developing more efficient computational algorithms, more accurate force fields and advanced sampling techniques [20–24], many biological phenomena are still not in the accessible range of AA approaches. Coarse-graining AA systems can significantly reduce degrees of freedom and accelerate system dynamics to model higher lengths and longer time scales. Chemical specificity preservative CG models such as MARTINI have been extensively employed to study the diffusion of lipid molecules [25], vesicle fusion, lipid rafting, [26,27] transmembrane protein aggregation, [28,29] and pore formation. [30] To further push the accessible length and time scale, various supra coarse-grained models of lipid molecules in which only a few beads represent the whole lipid molecule have been developed. [31–36] In explicit solvent approaches, the solvent particles are modeled explicitly

[a.] Department of Mechanical Engineering and Mechanics, Lehigh University, Bethlehem, Pennsylvania 18015, USA..
[b.] Department of Developmental Neurobiology, St. Jude Children's Research Hospital, Memphis, TN 38105, USA.
[c.] Department of Bioengineering, Lehigh University, Bethlehem, Pennsylvania 18015, USA †

# PAPER

and the dynamics of systems are usually modeled by molecular dynamics (MD) or dissipative particle dynamics (DPD). [37]

In the explicit solvent methods, majority of the simulation time is devoted to the calculation of interactions among solvent particles. Some implicit solvent approaches are efficient at the cost of a reduction in accuracy due to neglecting solvent degrees of freedom. Thus, the implicit solvent model of Cooke and Deserno [33] has been widely used in the simulation of membrane bending rigidity [38], protein aggregation [39], pore formation, [38,40] and membrane fission [41] due to its efficiency with high accuracy compared to other implicit solvent approaches. The implicit solvent CG models of Cooke and Deserno [33] can be coupled with fluid flow domains thermalized with stochastic fluctuations to consider the hydrodynamics effects without modeling all the solvent degrees of freedom explicitly. For instance, Atzberger et al. [42] and Wang et al. [43] developed an immersed boundary fluctuating hydrodynamics method that can be coupled with an implicit solvent CG model. Sterpone et al. [44] developed a multiscale model with a CG protein model that was coupled with a lattice-Boltzmann [45,46] flow solver. In a similar approach, Brandner et al. [47] and Yu and Dutt [48] coupled the fluctuating lattice Boltzmann method [46] with the implicit solvent Dry MARTINI [26] approach to study the self-assembly and aggregation of lipid bilayers and deformations of lipid vesicles under high shear rate flow. In addition to particle-based approaches, continuum scale models such as boundary element methods have also been employed in modeling the deformation of lipid vesicles in fluid flow.[49,50] Helfrich's free energy model [51] is usually used to find the bilayer response to hydrodynamic perturbations. However, such continuum scale models lack enough details to model the formation of pores or rupture and typically are only accurate at low deformation regimes.

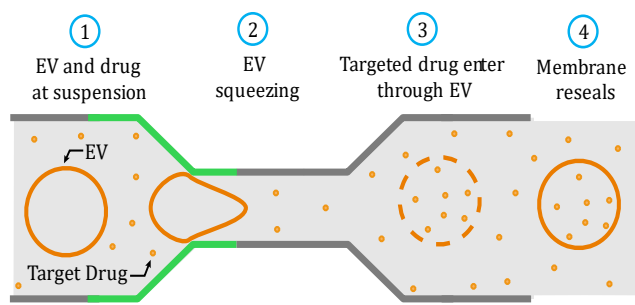

Figure 1: Drug loading through EV

In this study, coupling between a minimal CG model of a lipid bilayer with the fluctuating lattice-Boltzmann method is presented. The head-tail, implicit solvent, CG model of Cooke and Deserno [33] was used to represent the lipid molecules by only one head and two tail particles. The CG particles were coupled with a thermalized lattice-Boltzmann fluid flow solver.

Moreover, hydrodynamics was modeled by a computationally efficient, grid-based, lattice-Boltzmann method with good parallelization and scalability. By employing such a CG model and fluid flow solver, we could model the squeezing process of EVs with a diameter of tens of nanometers under realistic flow rates in a large computational box of a few hundred nanometers. The developed model was used to study the deformation of EVs due to squeezing, pore formation, and drug loading through the transient pores. Unlike the stain-based pore formation model [52], the pore formation process is simulated explicitly through coarse grained molecular dynamics coupled with a Lattice-Boltzmann flow solver, yet with large dimensional and time scales far beyond the reach of traditional molecular dynamics simulations. A novel approach for predicting pore formation and drug loading through the EV is presented by studying the effects of various squeezing parameters. This parametric study will provide guidance for the development of better EV drug loading conditions and devices.

## METHODS

### Simulation setup and parameters.

In this study, we employed the implicit solvent CG model of Cooke and Deserno [53] to model the lipid vesicles where one hydrophilic head and two hydrophobic tail particles represent the lipid molecule Figure 2 (c). Weeks-Chandler-Anderson potential given by Eq. 1 governed all the interactions among particles.

$$v(r) = \begin{cases} 4\varepsilon\left[\left(\frac{\sigma}{r}\right)^{12} - \left(\frac{\sigma}{r}\right)^{6} + 0.25\right] & r \leq r_c \\ 0 & r \geq r_c \end{cases} \quad (1)$$

In which the $\varepsilon$ and $\sigma$ are the energy and length units of the system respectively. Here, r is the distance between two particles and $r_c$ is the critical distance which is $r_c = 2^{1.2}\sigma$. We set $\sigma_{H-H} = \sigma_{H-T} = .95\sigma$ for the interactions of head particles and $\sigma_{H-H} = \sigma$ to ensure the cylindrical shape of lipid molecules. To represent the solvent effects, an attractive potential with the form of Eq. 2 with the range of $w_c$ is added only to the tail particles.

$$v_{att}(r) = \begin{cases} -\varepsilon & r < r_c \\ -\varepsilon \cos^2\left(\frac{\pi(r - r_c)}{2w_c}\right) & r_c \leq r \leq r_c + w_c \\ 0 & r > r_c + w_c \end{cases} \quad ..(2)$$

The range of attractive interactions is tuned by the $w_c$ parameter. This tunable parameter can affect the self-assembly and phase behavior of the lipid membrane and is set to $w_c = 1.6\ \sigma$ in the rest of this study [53–55]. Head and tail CG beads are bonded to each other with the finite extensible non-linear elastic (FENE) bond in the form of Eq. 3:

# PAPER

$$v_{bond}(r) = -\frac{1}{2}k_b r_0^2 \ln\left(1-\left(\frac{r}{r_0}\right)^2\right) \quad (3)$$

In which the bond constant is $k_b = 30\ \epsilon/\sigma^2$ and the maximum bond length $r_0 = 1.5\sigma$. The lipid molecule is straightened by a harmonic angular potential of the form Eq. 4.

$$v_{bend}(\theta) = \frac{1}{2}k_a(\theta - \pi)^2 \quad (4)$$

In which the angle constant is $k_a = 10\ \epsilon/rad^2$ and $\theta$ is the angle in radian made by the three CG beads of each lipid molecule. The above equation calculates the potential energy associated with the bond angle. This potential energy function is a harmonic function, meaning that the potential energy increases quadratically as the bond angle deviates from the equilibrium angle. The equilibrium angle between CG beads is considered $\pi$ radian in our case. The hydrodynamic is solved by the lattice-Boltzmann equations [56,57].

A stochastic fluctuation term that conserves the local mass and momentum is added to the stress tensor to address critical thermal fluctuations at the nanoscale. Details of the stochastic term implementation can be found in the works of Ladd [58] and Dunweg and Ladd.[59] We used the same grid size as the length unit of the CG model. Moreover, the time step is very similar to lattice-Boltzmann and MD simulations. The fluid domain is then coupled with the CG beads by a friction-based approach based on the implementation of Alrichs and Dunweg [46] in which the velocities of the fluid domain are interpolated for each CG bead and the thermalized coupling force can be written as Eq. 5.

$$\vec{F} = -\gamma(\vec{u} - \vec{v}) + \vec{\chi} \quad (5)$$

The coupling is tuned by the friction parameter $\zeta$. Here, $\vec{u}$ and $\vec{v}$ are the fluid and particle velocities respectively. $\vec{\chi}$ is a zero-mean stochastic force that satisfies Eq. 6:

$$\langle \chi_i(t)\chi_j(t')\rangle = 2\zeta k_B T \delta_{ij}\delta(t-t') \quad (6)$$

In which the $k_B$ is the Boltzmann constant, $T$ is the system temperature and $\delta$ is the Kronecker delta. Each particle is assumed to have the same mass that makes the unit of time

$$\tau = \sqrt{\frac{m\sigma^2}{\epsilon}}$$

reasonable. In this work, the simulation is carried out with a time step of $\delta t = 0.01\tau$ to ensure that the simulation accurately captures the dynamics of the lipid molecules and avoids numerical instability while keeping the simulation computationally inexpensive. We aim to develop a standard algorithm to model EV drug loading for different lipid conformations. Thus, the particles in our system do not correspond to any specific molecules, making the conversion to SI unit requires some interpretation. Each lipid tail can be considered a crude approximation of two-tail lipids, where each tail bead represents strands of five $CH_2$ groups. This assumption makes the length scale of the model $\sigma = .6\ nm$ and the mass scale of $N_A m \approx 140\ g/mol$. The temperature of the system is the same as body temperature, T = 310 K causes the units of energy being $\epsilon = 4.27 \times 10^{-21} J$.

The number of beads used for creating the EV varies between 44k~90k with a change of diameter, $60\sigma$ ~$80\sigma$. The area per lipid for outer and inner leaflet is measured by dividing the surface area of the leaflet by the lipid head number for each leaflet. The equilibrated EV has area per lipid (APL) at outer leaflet a⁺ and inner leaflet a⁻ $1.1\sigma^2$ and $1.3\sigma^2$ respectively, which is very similar to Cooke and Deserno[55]. The equilibrium APL in the outer and inner leaflet is different and the average APL is calculated by averaging the values. The average APL for the EV is measured to be $a_{avg} = 1.2\sigma$. Choice of potential width $w_c$ and temperature $K_B T$ are critical for defining the fluidity of the membrane. We choose $w_c$ and $K_B T$ value to be 1.6 and 1 respectively that to be in the fluidic phase to ensure in plane fluidity of the membrane. With the estimates of $\sigma = 0.6\ nm$ the calculated bilayer thickness is ~2.64 nm. The measured value is less than the typical bilayer thickness of ~5nm, but it is still in the lower but acceptable range of the reported range of bilayer thickness[60].

The EV is created using an in-house script, maintaining equal distance between head beads and aligning the tails with the head of the lipid as visualized in Figure 2 (a) and (b). Then the structure is equilibrated to get a stable stress-free configuration using Langevin thermostat with the temperature of $K_B T = 1\epsilon$ and timestep of $0.01\ \tau$ for $500\tau$ (see figure S1 for detail). The stress-free structure is verified by observing minimal potential energy with a fluctuation of < 1%. Then the EV is exposed to the fluid flow created through an implicit representation of fluid via the Lattice-Boltzmann Method (LBM)

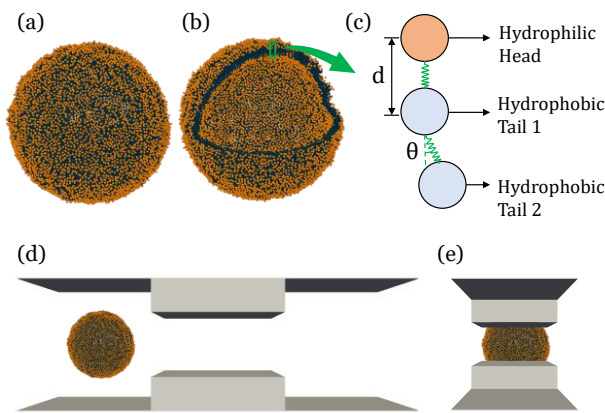

Figure 2: Simulation setup (a) XY view, (b) orthogonal view, (c) CG model of EV, (d) and (e) simulation box with EV

**PAPER**

of the ESPResSo package.[61] LBM is the first and most efficient

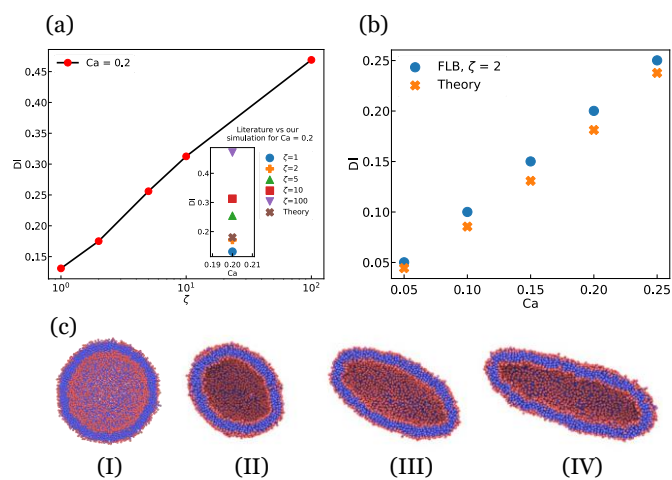

Figure 3: Comparison of our simulation result with literature[59]. (a) Variation of DI with various friction coefficients(ζ). ζ=2 shows agreement with Seifert et al[59]. (b) Variation of DI with various Ca. Calculated value shows good agreement with literature

lattice-based method that can be coupled with molecular dynamics, enabling the inclusion of hydrodynamic interactions into the simulation. In this method, Poiseuille flow is created by applying a homogeneous external body force density to the fluid in the MD unit. The simulation setup is visualized in Figure 2 (d) and (e).

The flow in a nanochannel around the EV is laminar. No-slip box boundary condition is used to simulate fluid flow throughout channel geometry. This boundary condition assumes that the fluid in contact with the wall is stationary and has zero velocity. Reynolds number is maximum at the narrow channel where the fluid has maximum velocity. The Reynolds number (Re) is calculated as $Re = \frac{v\rho L}{\mu}$. Here L is the characteristic length of the channel, $v$ is the velocity of the fluid, $\rho$ the density of the fluid, and $\eta$ the dynamic viscosity of the fluid. In our simulation, the characteristic or cord length of the largest rectangular channel is 160 $\sigma$. Using the value of dynamic viscosity $\eta = 1.98 \pm 0.16\ \sigma^{-2}\sqrt{m\epsilon}$, $\rho = 0.8\sigma^3$ and highest velocity $v \approx 1.0\sqrt{\epsilon/m}$ maximum Reynolds number calculated as Re = 564.

**Model agreement with literature.**

We analyzed the behavior of the lipid vesicle under a high shear rate flow. To verify the correct range of friction parameters, we compare the deformation index (DI) in our model with theoretical analysis of Seifert [62]. DI is defined as $DI = (l-s)/(l+s)$ in which the l and s are the lengths of the major and minor axis respectively. The Capillary number (Ca) can be defined as the ratio of the viscous drag to the surface tension, can be written in the form of $Ca = \mu \dot{\gamma} D^3/8E_b$, in which $\mu$ is the fluid viscosity, $\dot{\gamma}$ is the shear rate defined as $v_{\text{wall}}/H$ in which the H is channel height, $v_{\text{wall}}$ is the velocity of moving boundary condition, $E_b$ and is the bending rigidity which is

$E_b \approx 12\ k_B T$ [33,38]. The diameter of the vesicle is set to $D = 60\sigma$. To analyze the effect of friction coefficient on the DI of the vesicle, we set the Ca=0.2 and measured the DI after a long time equilibration of (to guarantee the stable shape of vesicle) under shear flow with various friction coefficients.

Figure 3a shows the effects of friction parameters on the DI of vesicles under shear flow. The vesicle may experience more deformations at the same Ca number at the higher friction coefficients. Figure 3c Shows the snapshots of vesicles under shear at Ca=0.2 and various friction parameters. As shown in Figure 3a, a good agreement with the theoretical results can be seen at $\zeta \approx 2$. We repeated the simulations with the friction parameter of $\zeta \approx 2$ at various Ca numbers from 0.05 to 0.25. As shown in Figure 3b, a good agreement with theoretical results can also be seen for Ca numbers less than 0.2. Thus, $\zeta \approx 2$ was chosen for the simulations in the rest of this paper.

**Pore area and drug diffusion analysis.**

The drug loading largely depends on the formation of pores on the EV surface, which needs to be accurately measured. However, it is difficult to calculate the size and area of the many pores formed on the EV since these pores are voids enclosed by discrete CG beads. To quantitatively measure the area of many pores formed on the EV surface, we converted the bead representation of the EV into a surface representation to analyze the pore formation. A triangulated surface mesh construction method developed by Stukowski [63] is followed to convert the three-dimensional particle into a geometric surface. This method uses the Delaunay tessellation method to construct the surface based on particle coordinates. It tessellates the three-dimensional spaces into tetrahedral simplices, which are subsequently classified as either belonging to an empty or filled spatial region. Finally, it constructs the surface manifold that separates the empty and filled region of space. To assign each tetrahedral Delaunay element to either the filled or the empty region concept of a probe sphere with a suitable radius is implemented. The empty region is a collection of all spaces where the probe can fit without touching any particles. The remaining tetrahedral regions are considered filled regions. Thus, the value of this probe radius is critical to calculating the empty region, and the probe sphere radius selection needs some investigation. It depends on how much atomic detail we want to capture in the pore definition. A smaller probe radius enables the capture of more detailed features with the cost of computational power. In our system, we choose the probe radius to be 1 nm which is the same value as the first nearest neighbor found by calculating $g(r)$ value (see Figure S4 for surface representation of EV using Delaunay tessellation method).

Sharei et al. [64] observed that the drug loading process happens within the first minute of the treatment process. This reported time scale for this process is independent of the

## PAPER

empirical parameters and system design. Also, according to McNeil and Steinhardt [65,66], it takes around 30 sec for the membrane to recover after transient pore opening. Previous work on pore dynamics suggested a longer pore opening time for largely deformed membranes under high-speed stretching. Therefore, assuming that the EV reaches maximum porosity during squeezing is reasonable. Based on experimental observation, we believe all pores will be healed after 30 sec following an exponential decay function regardless of their shape and size. The exponential decay function can be defined as:

$$A_{pore} = A_{max}^o \exp(-\gamma t) \quad (7)$$

Here $A_{max}^o$ is maximum pore area calculated right after squeezing, $\gamma$ is the decay constant and can be measured by considering $\frac{A_{pore}}{A_{max}^o} = 10^{-6}$ or a negligible value after 30 sec of squeezing. Drug loading inside the membrane is a diffusion process therefore, drug concentration inside the EV can be governed as:

$$\frac{\delta C}{\delta t} = \nabla \cdot (D_{eff} \nabla C) \quad (8)$$

Where $D_{eff}$ and $C$ is the effective diffusion coefficient converted into a surface integral using the divergence theorem. A surface integral on the EV surface can be approximated as:

$$\frac{\delta C}{\delta t} V_{exosome} = D_{eff}[(\nabla C)_{out} - (\nabla C)_{in}] A_{pore} \quad (9)$$

Here $(\nabla C)_{out}$ can be considered as zero as the concentration of the buffer can is higher than the concentration inside the EV. Considering drug concentration $C_b$ to be constant over time makes equation 9 further simplified.

$$-(\nabla C)_{in} = \frac{C_b - C}{L(\epsilon_A)} \quad (10)$$

Where L is the membrane thickness of EVs determined as a function of areal strain. More details on the calculation of membrane thickness can be found in our previous article[67]. By substituting equation 10 into equation 9 and solving the ordinary differential equation (ODE) with an initial condition C=0 at t=0 we get

$$\frac{C}{C_b} = 1 - \exp\left[\frac{D_{eff} A^o}{\gamma V_{exosome} L(\epsilon_A)} (\exp(-\gamma t) - 1)\right] \quad (11)$$

## RESULTS AND DISCUSSION

**Effect of flow velocity on EV squeezing and drug loading.**

The effect of squeezing velocity on drug loading is analyzed first. The magnitude of force applied on the LB points is tuned to archive the desired maximum velocity in the center of the channel. Velocity distribution through the nanochannel is discussed on supplementary material figure S2. Three cases with different flow velocities at the center of the channel are considered: a low velocity of $V_1 = 300 \ mm/sec$, a medium of $V_2 = 500 \ mm/sec$ and a high velocity of $V_3 = 700 \ mm/sec$. The velocity range is physiologically meaningful and consistent

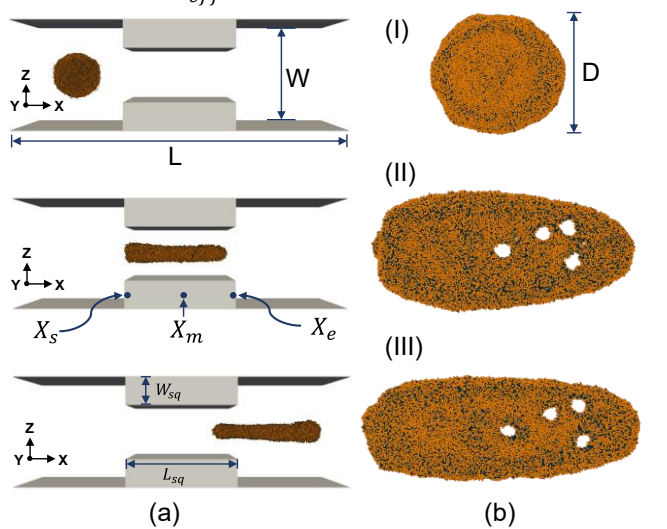

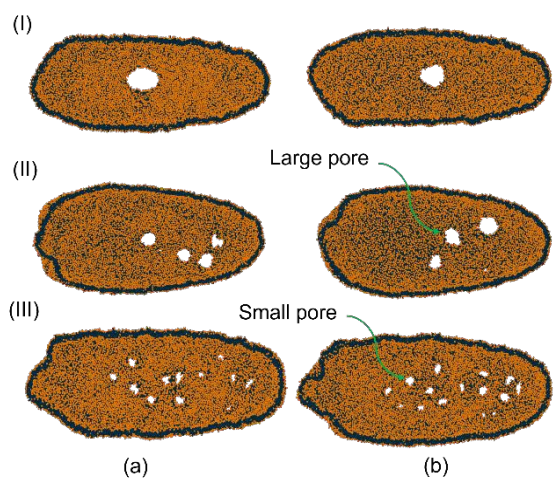

Figure 4: EV squeezing at a high flow velocity of 500 mm/sec. (a) Visualization of the simulation channel from XZ plane (b) molecular representation of EV from XY plane. (I) EV at entrance location $X_s$, (II) EV at middle of squeezing channel $X_m$, (III) EV at exit location $X_e$.

and drug concentration inside the EV respectively. The effective diffusion coefficient is considered to be $3 \times 10^{-15} \ m/s$ consistent with[64]. They obtained membrane diffusivity by simulating the diffusion of the delivery material into the cells suspended in a buffer of the delivery material. They also simulated the diffusion of material from the cell into a clean solution with no delivery material. The volume integral can be

Figure 5: Squeezing of EV under various flow velocities. (a) XY plane top view (b) XY plane bottom view; (I) $V_1$ = 300 mm/sec, (II) $V_2$ =500 mm/sec and (III) $V_3$ =700 mm/sec.

# PAPER

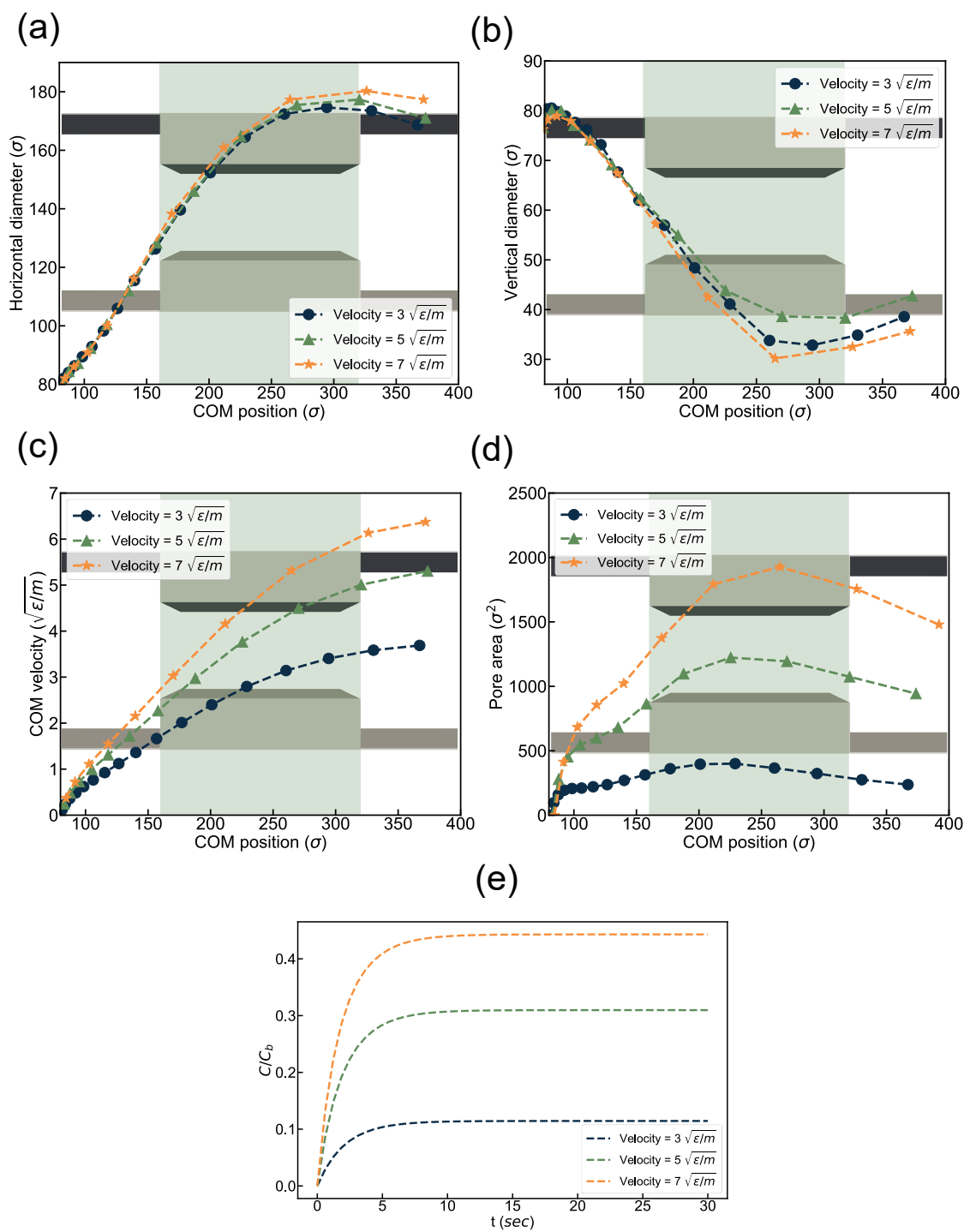

Figure 6: Simulation result under various flow velocities. (a) horizontal diameter, (b) vertical diameter, (c) Center of Mass (COM) velocity, and (d) Pore area vs. channel location (e) Drug loading over time.

with experimental results published by previous researchers [64,68]. For all cases, other parameters are kept constant to compare the effect of the flow velocity on the EV deformation. The channel length and width were fixed at L=6D and W=2D respectively, where D is the diameter of EV. The constriction

# PAPER

length $L_{sq}$ and constriction width $W_{sq}$ were 2D and 0.7D correspondingly.

Figure 4a shows the snapshot of EV at different simulation points for $V_2$ case, namely before the EV enter the squeezing channel ($X_s$), when the EV was at the middle of the constriction channel ($X_m$) and at the exit of the constriction channel ($X_e$). Figure 4.b visualizes the molecular representation of the EV for XY plane at the same time points indicated in Figure 4a. Figure 4b(II) and (III) visualize the pore opening due to the squeezing of the nanochannel.

Figure 5 visualizes the molecular representation of the pore opening for all the cases from the XY plane. It is visible from Figure 5 that with increasing velocity, the number of pores increases but the average pore radius decreases. Evans et al. [69] suggested that an increased loading rate between 1-10 mN/m for lipid vesicles results in higher critical rupture tension. Membrane rupture is prone to occur when the membrane tension increases above the critical. Higher flow velocity through the nanochannel results in a higher loading rate for the membrane, and the critical tension for pore opening increases. This phenomenon increases the internal energy of the membrane, and when internal energy crosses the critical limit, it generates an unstable pore in the membrane. With increasing critical internal energy value, energy stored before pore opening increases. This high internal energy reduces the area per pore but increases the number of pores generated. A similar trend is observed in our previous work[70], where an increased stain rate significantly increases total pore area but reduces average pore size. Under high strain rates, the membrane surface energy is released faster by generating lots of tiny pores rather than expanding existing pores. The complicated structure of the EV tail can be related to the difference in velocity between the EV tail center (located in the high-speed centerline area) and the tail edge (located near the wall). An increase in fluid velocity results in a more complicated tail structure as it increases the velocity difference between the center and edge of the tail. As squeezing velocity increases, deformed EV vertical and horizontal diameter change increases accordingly due to the positive relationship between the cell deformability and the fluid driving force. Pore recovery is not modeled here due to the larger timescale of the pore recovery process compared to the simulation time scale.

The instantaneous cell diameter along the vertical and horizontal direction is plotted in Figure 6(a) and (b) respectively. The constriction region for the nanochannel is shown as the shaded region in all the graphs. The initial position for all simulations is fixed at $X_s = 80\sigma$. The change in EV diameter is measured at the three zones described above. For all the cases, the horizontal diameter increases with the progression through the nanochannel.

In the beginning, the diameter changes at a lower rate, increasing up to the middle of the constriction. Then the EV diameter change becomes stable during squeezing. For the rest of the channel, diameter decreases again as part of the EV already passed the narrow constriction. Vertical diameter shows an inverse trend in comparison to horizontal diameter. It starts with a slight decrease with a higher decrease in diameter up to the middle of the constriction. Then it shows a stable diameter followed by an increase in diameter at the end of the channel. With increasing flow velocity, horizontal diameter increases to a greater extent and vertical diameter decreases at the same rate as the volume of the EV need to be constant. Figure 6 (c) visualizes the change of Center of Mass (COM) velocity at different channel locations.

We observed an increase in COM velocity with a maximum velocity at the center of the constriction for all cases. The EV velocity increases as the flow velocity increase at the center of the channel due to the constriction. The velocity profile observed here is consistent with the experimental value observed in Sharei et al..[64,68,71] Due to EV deformation, transient pores appear on the EV membrane. Figure 6 (d) visualizes the EV pore area at different channel locations. The porosity of the EV membrane increases until it reaches $X_m$. Subsequently, the porosity starts to decrease due to the recovery contraction of the cell membrane. The generation of transient pores shows a

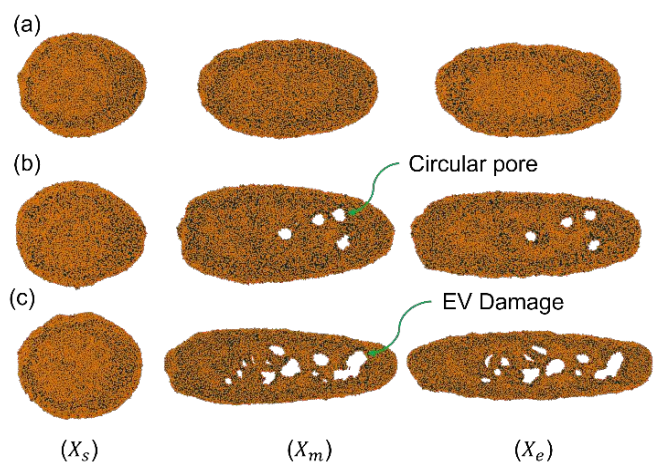

Figure 7: EV shape during squeezing at $V_2 \approx 500$ mm/sec and L=6D. (a) No pore under constriction width $W_{sq} = .8D\ \sigma$; (b) circular pore for constriction width $W_{sq} = .7D\ \sigma$; (c) EV damage for constriction width $W_{sq} = .6D\ \sigma$.

positive relationship with the flow velocity. With the increase in flow velocity, the pore area increases significantly. At the beginning of the EV constriction, the area increases slowly, reaching a maximum value at the center. Afterward, the area again starts to decrease due to EV shrinkage. Figure 6(e) represents the variation of drug loading rate within 30 sec of pore lifetime. The drug loading rate increases significantly with the flow velocity due to higher pore area generation. It is also visible from the graph that with the increase in flow velocity,

**PAPER**

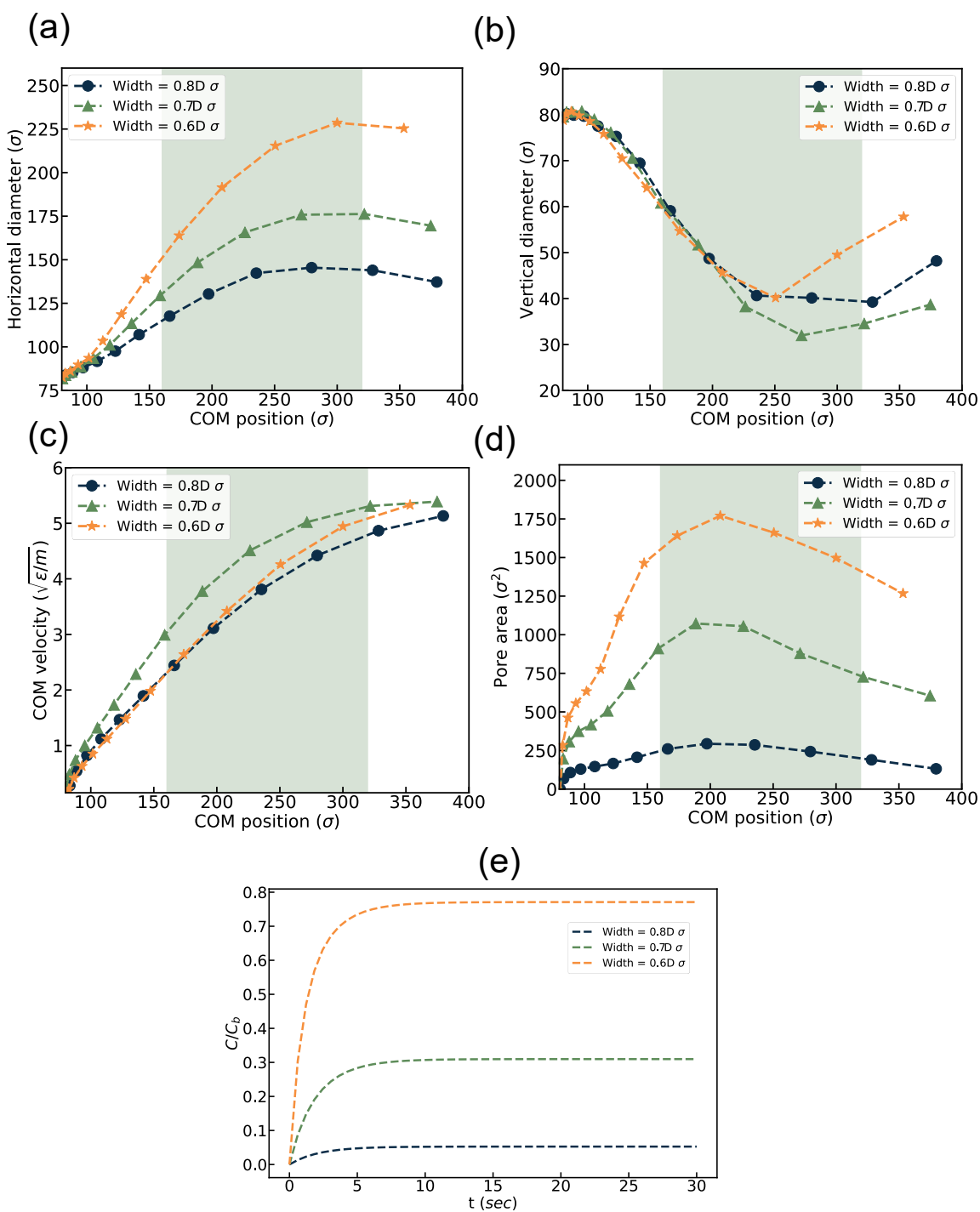

Figure 8: EV squeezing simulation result under various channel widths. (a) horizontal diameter, (b) vertical diameter, (c) Center of Mass (COM) velocity, (d) Pore area vs. channel location, (e) Drug loading over time.

the time required to reach the steady state also increases. This phenomenon is in agreement with the previous experimental results.[64,68] Increase in flow velocity from $V_1$ to $V_2$ bumped up drug loading by ∼200%, while an increase in velocity from $V_2$ to $V_3$ enhanced the drug loading by ∼40%

## PAPER

**Effect of channel width on EV squeezing and drug loading.**

We squeezed EVs at different constriction widths to study the effect of channel width on drug loading. We have considered three constriction widths, *i.e.*, $W_1 = 0.6D, W_2 = 0.7D, W_3 = 0.8D$, where D is the diameter of the EV. For all the cases, EV velocity and constriction length are kept constant at $V_2 \approx 500\ mm/sec$ and L=6D respectively. EV shapes at different channel locations are visualized in Figure 7. EV protrusion deformation is more apparent in narrower channels compared to the others. As we kept EV average velocity identical for all the cases, EVs required a greater driving force to go through the narrower channels. This phenomenon results in higher membrane stretch for narrower channels. The EV tail has a more complex tail shape for higher driving force through the narrower channels, which causes a higher cell elongation towards the flow direction. It is observed from Figure 7 that narrower constriction causes the generation of a larger number of pores as well as a larger pore area. EV cells expand along the flow direction until they cross the narrow constriction. Afterward, it starts to shrink and transient pores healing takes place. It is observed from Figure 7(c) that although a narrower constriction creates more pores in the EV, having an extremely narrower constriction can potentially damage the EV permanently and stop the EV from recovering to its original shape (See supplementary material for EV damage criteria). Thus, designing constriction width for squeezing EVs needs proper attention and the value needs to be optimal, facilitating desired drug loading while keeping the EV recoverable.

Figure 8 (a) represents the change in horizontal diameter as a function of COM location along the channel. In all graphs, the constriction regions are shaded zone. All the graphs indicate a similar trend, as discussed in previous section. It shows a similar trend of increase in horizontal diameter as it propagates through the constriction, which is consistent with Figure 7. With the increase in channel constriction, a decrease in the EV diameter becomes more prominent as it faces more driving force to go through the constriction. Figure 8 (b) shows that as the horizontal diameter increases, the vertical diameter decreases as EVs move along the channel due to the conservation of volume. EV velocity as a function of COM location for different constriction widths is illustrated in Figure 8 (c) EV drug loading increases at a higher rate for reduced constriction than in the other cases due to higher flow velocity. However, the difference in COM velocity for various constrictions is minimal as the fluid inlet velocity for the channel was kept constant for all cases. Figure 8(d) represents the change in pore area for different cases of channel constriction. As observed in Figure 7 the number of pores generated in narrower constriction is higher than in others. This occurrence results in a larger pore area for narrower constriction, as shown in Figure 8 (d). The normalized drug concentration inside the EV cell is visualized in Figure 8 (e). Drug concentration has a positive relationship with porosity and an increase in porosity results in higher drug concentration inside the EV. This trend is consistent with previously published results from Sharei *et. al.*.[64,68] It should be mentioned that decreasing constriction from 0.7D to 0.6D results in more drug concentration in the EV compared to the change from 0.8D to 0.7D. Reducing constriction from 0.8D to 0.7D results in a ~500% increase in drug concentration, whereas a change in constriction from 0.7D to 0.6D causes a ~166% increase.

**Effect of EV diameter on EV squeezing and drug loading.**

This section analyzes the effect of EV diameter on pore formation and drug loading. We have simulated three cases with different EV diameter $D_1 = 60\sigma, D_2 = 70\sigma$ and $D_3 = 80\sigma$. All the channel length is kept constant at 6D to keep the EV diameter to channel geometry ratio fixed. Here D is the diameter of the EV. For each case, the EV is positioned at X = 1D location. The constriction length of 2D is considered in the middle of the channel. All the other parameters like flow velocity are kept constant at $V_2 \approx 500\ mm/sec$ and L=6D respectively. Cell shapes at different channel locations are illustrated in Figure 9. As shown in the Figure 9 EV with $60\sigma$ diameter does not have any visible pore whereas $70\sigma$ diameter EV showed only 3 nanopores on the top view of the membrane. On the other hand, $80\sigma$ diameter EV shows 4 nanopores, and pore areas are more prominent than in all other cases. The increase in EV diameter shows a more complex tail geometry as it has a higher membrane area that deforms easily compared to the EV with a lower membrane area.

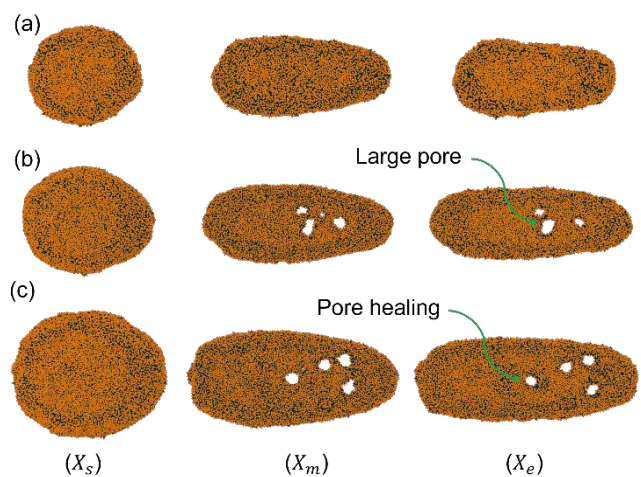

Figure 9: EV shape during squeezing at $V_2 \approx 500\ mm/sec$ and L=6D for various EV sizes. (a) EV diameter $D = 60\ \sigma$, (b) EV diameter $D = 70\ \sigma$, (c) EV diameter $D = 80\ \sigma$

PAPER

The horizontal diameter of EV at different COM positions is visualized in Figure 10(a). Each graph starts from a different starting point as we positioned them at X=1D to keep the diameter to channel length ratio constant. Each case has a constriction region of 2D shaded with different colors on all graphs. EV with a larger diameter has a larger expansion

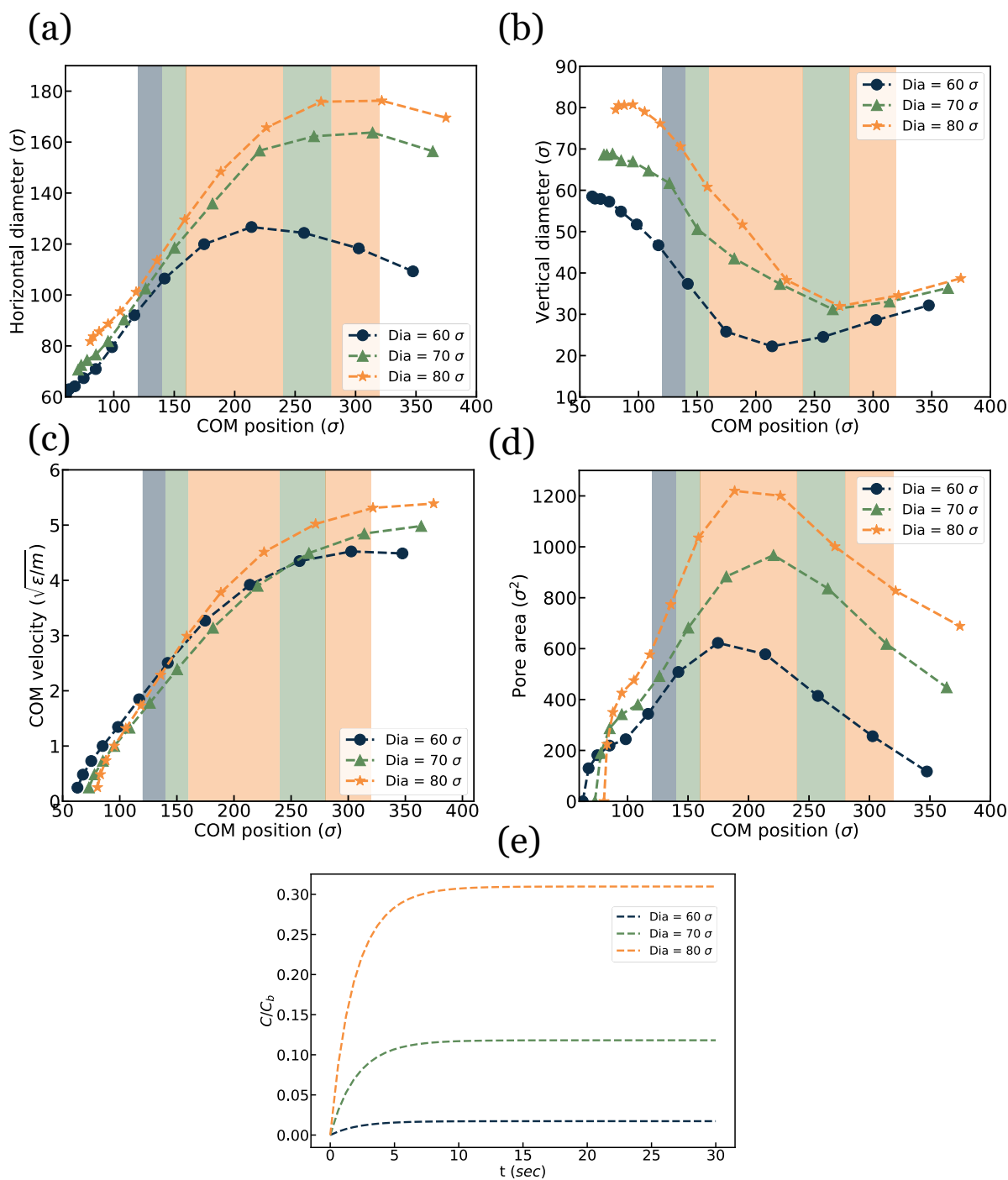

Figure 10: EV squeezing simulation result for EV of various diameters. (a) horizontal diameter, (b) vertical diameter, (c) Center of Mass (COM) velocity, (d) Pore area vs. channel location, and (e) Drug loading over time.

# PAPER

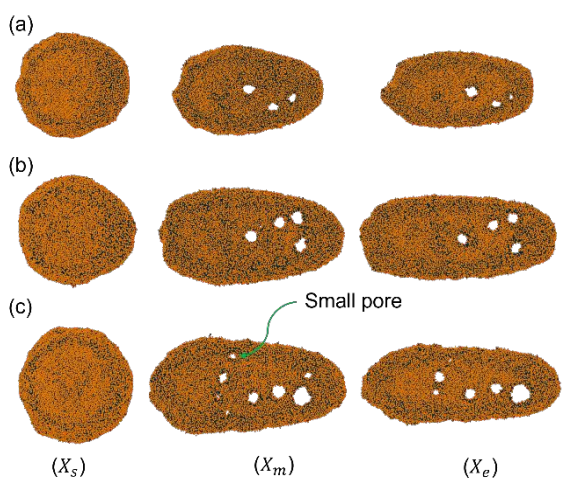

Figure 11: EV shape and pore formation during squeezing at : $V \approx 500\ mm/sec$ and $W_{sq}$=0.7D for various channel lengths (a) Very few pores for constriction length $L_{sq}$ =1D, (b) medium size pores for constriction length $L_{sq}$ =2D, (c) small and large pore for constriction length $L_{sq}$ =3D.

deformation than the other cases. Figure 10(b) illustrates the decrease in vertical diameter as a function of COM position. The EV's vertical diameter development decreases as it passes through the narrow constriction due to the conservation of volume. Figure 10 (c) visualizes the change in COM velocity for different EV diameters.

The pore area for EVs of different diameters is presented in Figure 10(d). It is observed that the EV with a larger diameter has a larger pore area compared to the smaller diameter counterparts. It should be mentioned that the EV with $D_1 = 60\sigma$ in Figure 9 shows no pore in molecular representation. However, using the Delaunay tessellation method, surface representation showed some transient pore and the pore area for drug loading is calculated based on that post analysis. The normalized drug concentration inside the EV cell is visualized in Figure 10(e). Drug concentration has a positive relation between porosity and an increase in porosity resulting in higher drug concentration inside the EV

**Effect of channel length on EV squeezing and drug loading.**

Lastly, we checked the effect of squeezing length/duration on EV drug loading by varying constriction length from $L_{sq_1} = 1D, L_{sq_2} = 2D$ and $L_{sq_3} = 3D$. The channel length is chosen as

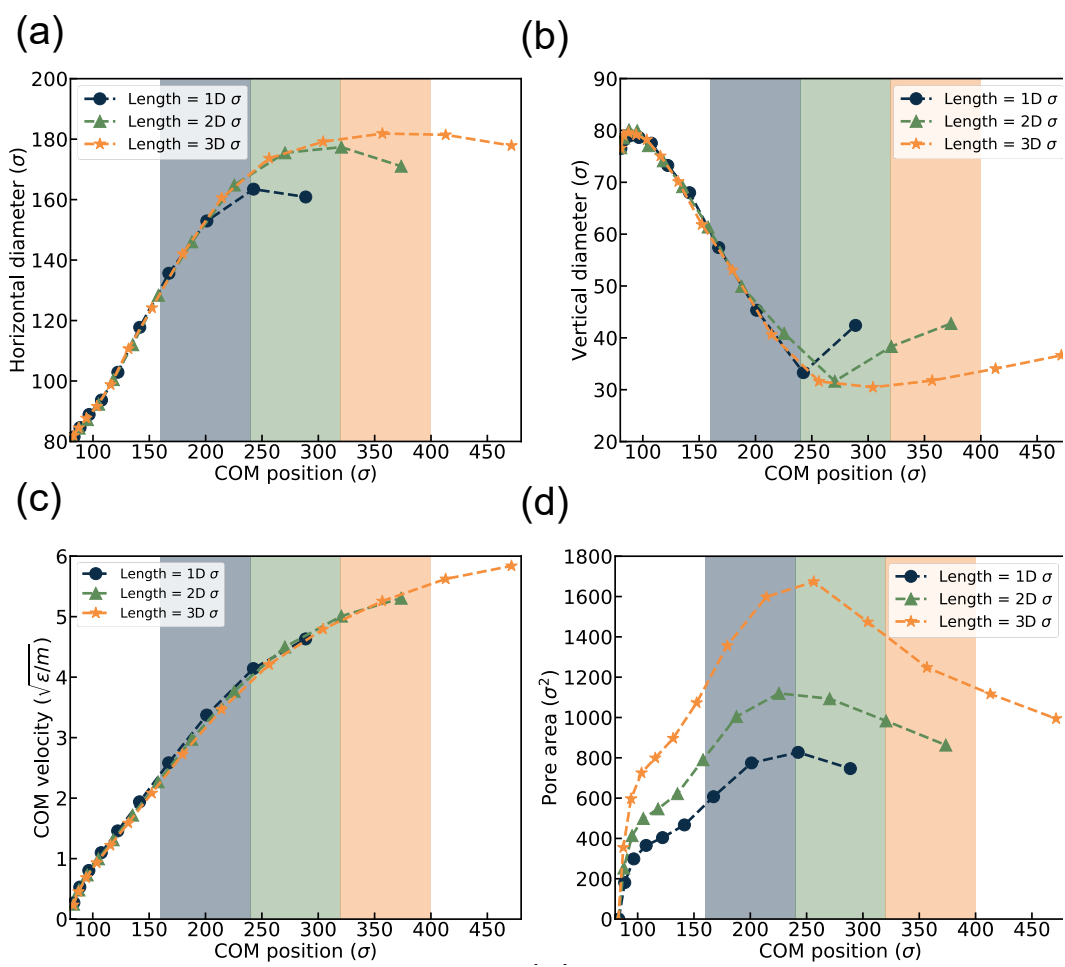

**PAPER**

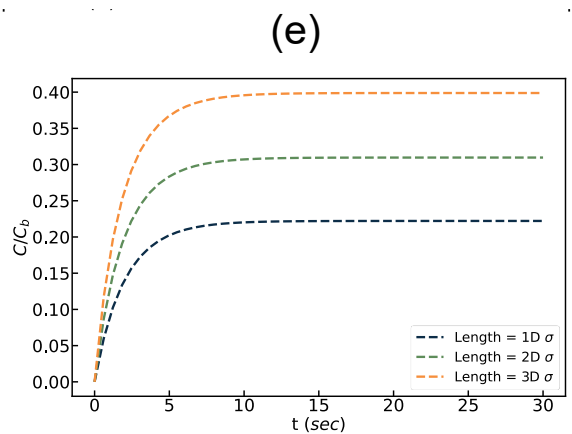

Figure 12: EV squeezing simulation results under various channel lengths. (a) horizontal diameter, (b) vertical diameter, (c) Center of Mass (COM) velocity, and (d) Pore area vs. channel location (e) Drug loading over time.

$L_1 = 5D, L_2 = 6D$ and $L_3 = 7D$ to keep the inlet and outlet distance constant for all cases. All the other parameters like flow velocity and channel width are kept constant at $V_2 \approx 500\ mm/sec$ and W=2D respectively. Cell shapes at different channel locations are illustrated in Figure 11. The cell dynamics and deformation are pretty similar to that described in the previous sections. As shown in Error! Reference source not found., the EV deformation is larger for greater constriction lengths.

Figure 12(a) shows the change in EV diameter as a function of COM position. The EV diameter changes with increased constriction length. In Figure 12 (b), we observe an opposite relationship between the vertical diameter and constriction length consistent with the abovementioned results. EV COM velocity as a function of COM position is visualized in Figure 12 (c) where the trend is almost the same for all cases as all the flow is already developed and changing channel length does not change flow condition.

As shown in Figure 12 (d) pore area increases as the constriction length increases and is consistent with the result above. Finally, Figure 12 (e) shows the normalized drug concentration inside the EV with respect to loading time. It suggests that drug loading is a function of channel length; a longer channel leads to higher normalized drug loading. Therefore, the result is consistent with experimental findings [64,68] suggesting that higher loaded drug concentration can be achieved by increased squeezing channel length.

To provide guidance on how to design proper channel geometry and EV velocity for effective pore opening and drug loading, we performed simulations under a large range of EV squeezing velocities and nanochannel widths. In Figure 13, a pore opening status phase diagram is generated as a function of EV velocity and constriction width. We categorize EV squeezing results into damage zone, safe zone, and no pore zone. Damage zone is where permanent damage or rupture happens on EV. No pore zone is where there is no pore formation on EV. Safe zone is the suggested status where transient pores are formed on EV enabling drug loading. From the phase diagram, it is clear that lower constriction width and higher exoso EV me velocity can potentially damage the EV, based on the criteria described in the supplementary material (Figure S3). On the other hand, squeezing the EV with a higher width and lower EV velocity

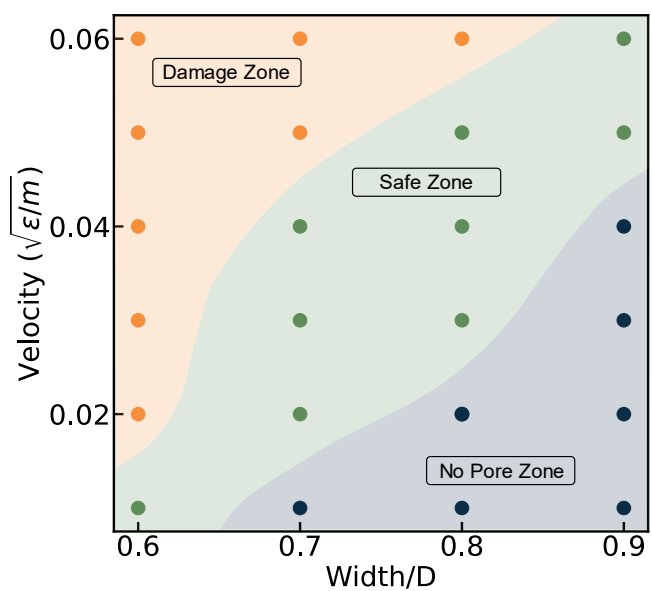

Figure 13: Phase diagram of pore formation status under different EV velocities and constriction widths. Damage zone is where permanent damage or rupture happens on EV. No pore zone is where there is no pore formation on EV. Safe zone is the suggested status where transient pores are formed on EV enabling drug loading.

# PAPER

results in no pore creation, thus no drug loading into the EV. In between the above two zones is a safe zone that generate transient pores on EV which is favorable for drug loading.

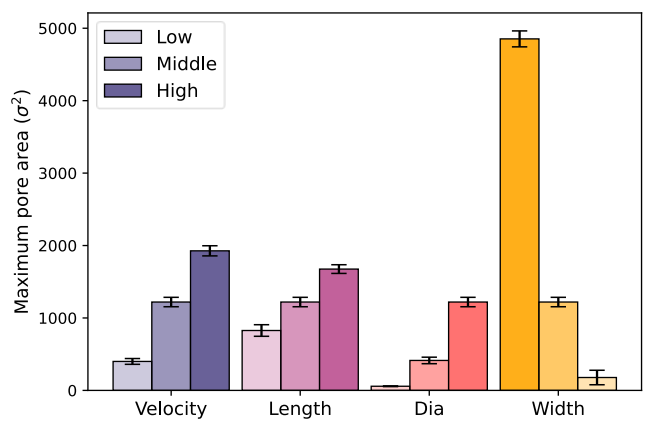

Figure 14: Summary of maximum pore area for EV squeezing under various conditions.

Figure 14: Summary of maximum pore area for EV squeezing under various conditions. summarizes the maximum pore area for various cases presented above. The pore area under each parameter's lower, middle and higher value results were reported. The lower value for velocity is $300\ mm/sec$, with a median value of $500\ mm/sec$ and a higher value of $700\ mm/sec$. Similarly, for the length case, the constriction length values are 1D, 2D and 3D respectively. For the diameter case, the diameters are considered for $60\sigma, 70\sigma$ and $80\sigma$ respectively. Lastly, for the width case, the constriction widths of 0.6D, 0.7D and 0.8D are considered respectively. The lower constriction width case shows an abnormally high maximum pore area. The observation is consistent with Figure 7 and results from the highly narrow constriction that causes the EV to be damaged. Thus, the EV drug loading is very sensitive to channel width change, followed by a change in flow velocity.

## CONCLUSION

We have coupled a computationally inexpensive supra CG model with fluctuating lattice-Boltzmann method, which enables us to simulate the entire EV squeezing process with a reasonable computational cost. Each lipid molecule is represented by one head particle and two tail particles. In-house code is used to compute the pore formation and relate the pore area with drug loading through the diffusion process. The proposed method can predict the drug loading into the EV and thus can help tune the squeezing parameters for desired drug loading. The method is free from any empirical parameter depending on operating conditions and channel geometry. Our results show that drug loading through the EV increases with increasing flow velocity, increasing EV diameter, decreasing constriction width, and increasing constriction length. An optimal value for channel width can be predicted to achieve maximum drug loading without damaging the EV. EV squeezing through nanochannel is a complicated phenomenon in biophysics; thus, the coarse-grained simulation presented in this work can provide insights into both the physics during the dynamic process and how various parameters influence drug loading. EV squeezing velocity, constriction width and length, properties of the EV such as their sizes, shapes, and initial positions are parameters that can significantly change the loading results. Analytical equation used to calculate drug loading is one of the limitation of the proposed method that assume pores to be sealed after 30 s. However, pore closing time can differ depending on pore radius[70]. We are working on developing a more efficient model that can simulate the pore closing with reasonable computational cost. The proposed model can be used to optimize squeezing parameters like flow velocity, channel geometry parameter and EV diameter to obtain preferred drug loading through EV.

## Conflict of Interest:

The authors declare no competing financial interest

## Supplementary Information available:

See the supplementary material for the EV damage criteria

## Authors' Contributions:

K.I. designed research, performed research, analyzed data, and wrote the paper. M.R. designed research and wrote the paper. Y.L. designed and directed research, performed research, analyzed data, and wrote the paper.

## Acknowledgments

This work was supported by National Institute of Health Grant R01HL131750, R21EB033102, National Science Foundation Grant CBET 2039310, EECS 2215789, Pennsylvania Department of Health Commonwealth Universal Research Enhancement Program (CURE), and Pennsylvania Infrastructure Technology Alliance (PITA).

## PAPER

PAPER

# Supplementary materials

## 1. EV initial structure:

Generating an initial stable EV structure for molecular dynamics simulations is crucial in accurately studying EVs and their behavior. While there are multiple methods for generating EV structures, one widely adopted approach involves utilizing an in-house script to create the EV and then equilibrating it to achieve stability.

The initial creation of the EVs involves carefully placing the lipid molecules in a periodic box and arranging them in a bilayer configuration while aligning the tails with the head of the lipid and maintaining a constant distance between the head groups. Following this, the EV is subjected to equilibration, which allows it to relax and attain a stable configuration. Equilibration is a fundamental step in the process, as it involves simulating the EV under conditions that allow for relaxation and equilibrium. We enable the initial structure to be equilibrated using a Langevin thermostat while closely monitoring the fluctuations in shape, size, and orientation of the lipids, and the simulation is run for a sufficient amount of time until a stable liposome configuration is obtained.

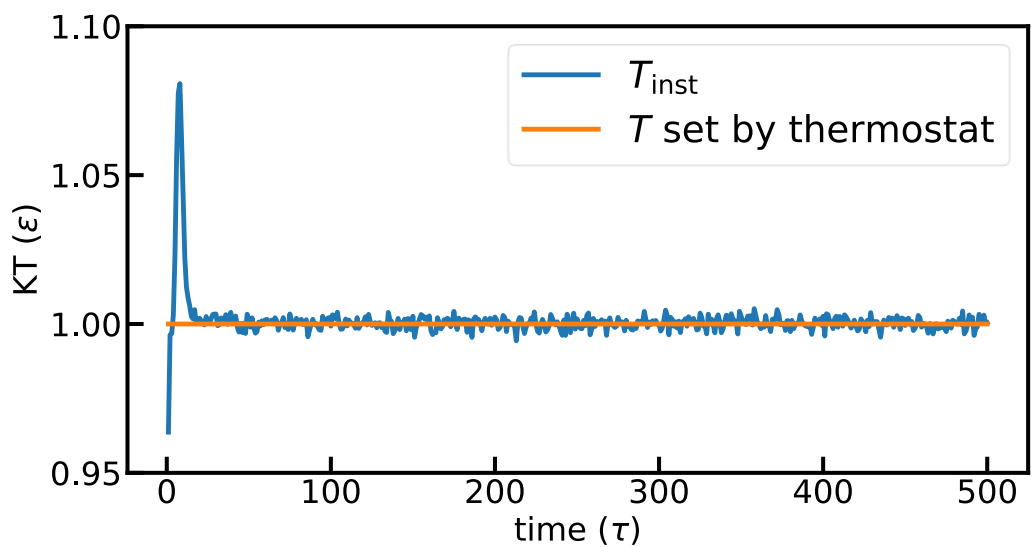

**Figure S1:** Temperature fluctuation over time

In Figure S1, we can observe the temperature fluctuations over time. The temperature visualization includes two distinct colors to differentiate between the instantaneous temperature and the temperature the thermostat defines. Blue represents the instantaneous temperature, while orange represents the temperature specified by the thermostat. The figure clearly shows that the temperature initially spiked up at the very beginning of the simulation. This spike can be attributed to the unstable initial structure manually created using our script. However, it's important to note that this spike was quickly brought down to the desired temperature by implementing the Langevin thermostat. The Langevin thermostat is a widely used method to control the temperature in molecular dynamics simulations by mimicking the effect of a heat bath on the system. With the Langevin thermostat, the temperature was tabilized and fluctuated around the set temperature. This fluctuation is natural, as expected, and indicates that the simulation is proceeding with a minimized stable configuration.

## 2. Velocity profile:

In our simulation, Poiseuille flow is generated by specifying uniform force per unit of volume. By carefully adjusting the applied force, we have achieved the desired maximum velocity at the center of the channel. Our simulation results are presented in Figure S2, which clearly visualizes the fluid velocity profile at various flow conditions. The velocity profile is obtained by employing a polynomial fit to the fluid velocity data along the x direction. Our graph illustrates the gradual increase of velocity along the constriction region, where it reaches a maximum value, followed by a gradual decrease to a minimum value in the remaining parts of the channel. This detailed representation of the velocity profile provides valuable insights into the Poiseuille flow behavior within the nanochannel.

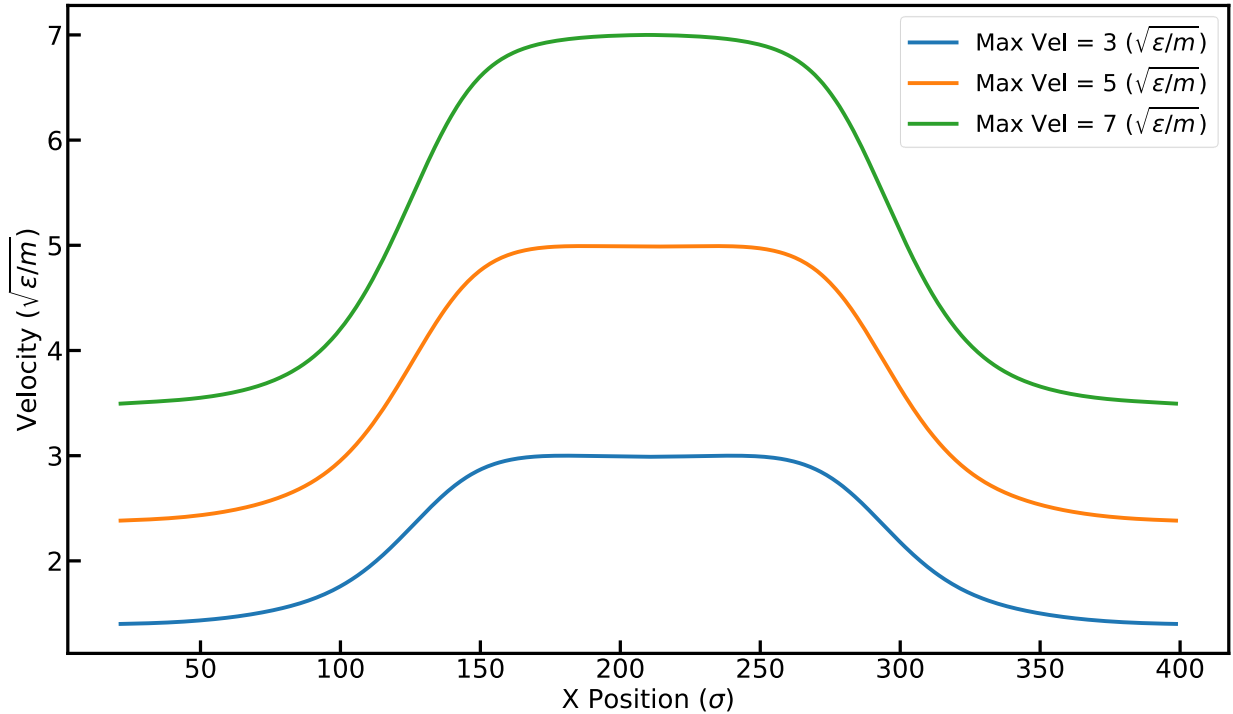

**Figure S2:** Fluid velocity profile over different regions of the nanochannel

## 3. EV damage criteria:

Extracellular vesicle (EV) being a subtype of extracellular vesicle and made of lipid-bilayer nanoscale membrane particles (one head and two tails), we believe the damage of EV should follow similar failure criteria as a standard biological membrane. Literature review [1–3] reveals that bilayer membrane ruptures are prone to occur between tension ~1-25 mN/m, which corresponds to strain value in the order of 2 -5%. Evans et al [4] found two distinct regimes for rupture: low-strength cavitation-limited and a high-strength defect-limited regime with a transition at loading rate(tension/time) for DOPC bilayer membrane around 10 mN/m/s. They show that membrane failure is not a static material property but a function of dynamic loading. Membrane tension can vary from 6 mN/m to 13 mN/m for loading rates of 0.07 mN/m/s and 25 mN/m/s respectively. Our squeezing simulation of EVs through nanochannel causes the EV to be damaged. Zevnik[5] et al. show through simulation that for a high loading rate in the range of ∼ $10^9 - 10^{10}$ mN/m/s critical rupture strength rises logarithmically and stays between 80 to 95 mN/m. Their results closely mimic the molecular dynamics simulation of liquid-phase DPPC bilayer[6]. Based on the values from the literature, we considered two criteria for EV damage. The primary damage criterion is the areal strain value of 0.203 and the secondary damage criterion is the aggregation of pores.

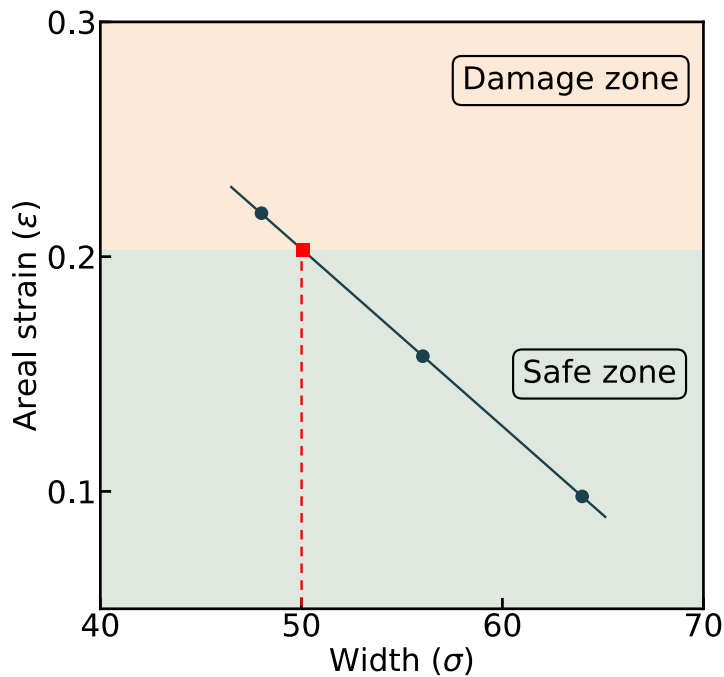

**Figure S3:** Safe vs damage zone of EV for change of width case

**Primary damage criteria:**

The primary cause of EV membrane damage is the creation of pores caused by squeezing the EV through the nanochannel. Zevnik et al. [5] show that for an equibiaxial loading case, the critical stress value is 20MPa. Although the pore formation and subsequent EV damage largely depend on the strength of the membrane here, we consider a critical areal strain of value 0.203 calculated from obtained linear strain by Zevnik et al. [5].

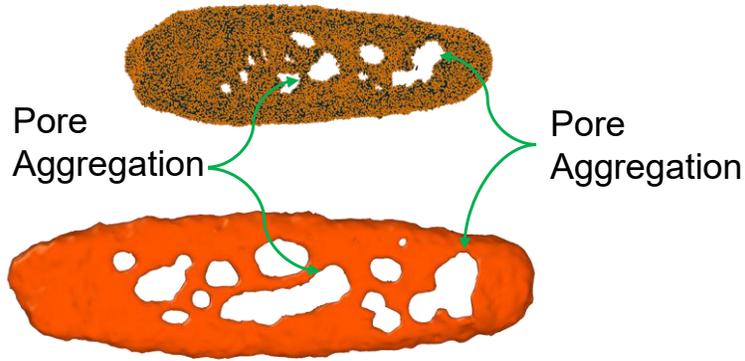

**Figure S4:** Pore aggregation for $W_1 = 0.6D$ case

**Secondary damage criteria:**

The secondary cause of EV damage is due to aggregation of pores. If two or more pore aggregate together, it creates a massive irregular-shaped pore in the membrane, potentially damaging the cell. The aggregation of pores is inspected after constructing the surface mesh, as described in section 2.2 of the paper. Most of the single pore is roundish in shape if no pore aggregation happens.

As shown in figure S1, the threshold areal strain value of .203 divides the graph into the damage zone and safe zone. The first case from the width change fulfils both criteria to be considered a damaged EV case. Figure S1 shows the areal strain for various constriction width cases. Areal strain for width $W_1 = 0.6D$ is calculated as 0.218, above our estimated threshold value of 0.203 and falls in the EV damage zone. A trend line is drawn connecting the three data points, and the intersecting point between the threshold value and the trendline is marked with a red box. The intersection point is then extrapolated towards the x-axis to get the desired minimal channel width value to achieve maximum exosomal pore without damaging the EV. Figure S2 shows the aggregation of pores with surface representation as described in section 2.2 of the paper. The pore aggregates as the exos EV ome propagate along the flow direction, creating a larger irregular-shaped pore.